\documentstyle[11pt,aaspp4]{article}

\def\lae{\mathrel{<\kern-1.0em\lower0.9ex\hbox{$\sim$}}}
\def\gae{\mathrel{>\kern-1.0em\lower0.9ex\hbox{$\sim$}}}
 
\slugcomment{Accepted for publication in the Astrophysical Journal Letters}
 
\lefthead{C\^ot\'e et al.}
\righthead{A Carbon Star in M14}
 
\begin{document}
 
\title{Discovery of a Probable CH Star in the Globular Cluster M14 and Implications
for the Evolution of Binaries in Clusters}
 
\author{Patrick C\^ot\'e,\altaffilmark{1,2,3} David A. Hanes,\altaffilmark{1,4} Dean E. McLaughlin,\altaffilmark{5} T.J. Bridges,\altaffilmark{1,6} James E. Hesser\altaffilmark{3} and Gretchen L. H. Harris\altaffilmark{7}}
 
% Notice that each of these authors has alternate affiliations, which
% are identified by the \altaffilmark after each name.  The actual alternate
% affiliation information is typeset in footnotes at the bottom of the
% first page, and the text itself is specified in \altaffiltext commands.
% There is a separate \altaffiltext for each alternate affiliation
% indicated above.
 
\altaffiltext{1}{Visiting Astronomer, Canada-France-Hawaii Telescope, operated by the
National Research Council of Canada, the Centre Nationale de la Recherche de France, and the University
of Hawaii.}
\altaffiltext{2}{Visiting Astronomer, WIYN Observatory. The WIYN Observatory 
is owned and operated by the WIYN Consortium, which consists of the University of Wisconsin, 
Indiana University, Yale University, and the National Optical Astronomy Observatories 
(NOAO).}
\altaffiltext{3}{Dominion Astrophysical Observatory,
Herzberg Institute of Astrophysics, National Research Council,
5071 West Saanich Road, Victoria, BC, V8X 4M6, Canada}
\altaffiltext{4}{Physics Department, Queen's University, Kingston, ON, K7L 3N6, Canada}
\altaffiltext{5}{Department of Physics \& Astronomy, McMaster University, Hamilton, ON, L8S 4M1, Canada}
\altaffiltext{6}{Royal Greenwich Observatory, Madingley Road, Cambridge, CBS 6EZ, England}
\altaffiltext{7}{Astronomy Program, Guelph-Waterloo Program for Graduate Work in Physics, Waterloo Campus,
Department of Physics, University of Waterloo, Waterloo, ON, N2L 3G1, Canada}
 
% The abstract environment prints out the receipt and acceptance dates
% if they are relevant for the journal style.  For the aasms style, they
% will print out as horizontal rules for the editorial staff to type
% on, so long as the author does not include \received and \accepted
% commands.  This should not be done, since \received and \accepted dates
% are not known to the author.
 
\clearpage
\begin{abstract}
We report the discovery of a probable CH star in the core of the Galactic globular cluster
M14 (= NGC 6402 = C1735-032), identified from an integrated-light spectrum of 
the cluster obtained with the MOS spectrograph on the
Canada-France-Hawaii telescope.
From a high-resolution echelle spectrum of the same star obtained with the 
Hydra fiber positioner and bench spectrograph on
the WIYN telescope, we measure a radial velocity of $-53.0\pm1.2$ km s$^{-1}$.
Although this velocity is inconsistent with published estimates of the systemic radial velocity of M14
(e.g., ${\bar {v_r}} \approx -123$ km s$^{-1}$), we use high-precision 
Hydra velocities for 20 stars in the central
2\farcm6 of M14 to calculate improved values for the cluster mean velocity and 
one-dimensional velocity dispersion: $-59.5\pm1.9$ km s$^{-1}$ and $8.2\pm1.4$ km s$^{-1}$, respectively.
Both the star's location near the tip of the red giant branch in the cluster color magnitude
diagram {\sl and} its radial velocity therefore argue for membership in M14.
Since the intermediate-resolution MOS spectrum shows not only enhanced CH absorption but also strong
Swan bands of C$_2$, M14 joins $\omega$~Cen as the only globular clusters known to contain
``classical" CH stars. 
Although evidence for its duplicity must await additional
radial velocity measurements, the CH star in M14 is probably, like all
{\sl field} CH stars, a spectroscopic binary with a degenerate (white dwarf)
secondary. The candidate and confirmed CH stars in M14 and $\omega$~Cen, and
in a number of Galactic dSph galaxies, may then owe their existence to the
long timescales for the shrinking and coalescence of hard binaries in
low-concentration environments.
\end{abstract}
 
% The different journals have different requirements for keywords.  The
% keywords.apj file, found on aas.org in the pubs/aastex-misc directory, 
% contains a list of keywords used with the ApJ and Letters.  These are 
% usually assigned by the editor, but authors may include them in their 
% manuscripts if they wish. 
 
\keywords{stars: carbon --- stars: binaries: --- Galaxy: stellar content --- 
globular clusters: individual (M14) --- galaxies: Local Group --- galaxies: stellar content}
 
% That's it for the front matter.  On to the main body of the paper.
% We'll only put in tutorial remarks at the beginning of each section
% so you can see entire sections together.
 
% In the first two sections, you should notice the use of the LaTeX \cite
% command to identify citations.  The citations are tied to the
% reference list via symbolic KEYs.  We have chosen the first three
% characters of the first author's name plus the last two numeral of the
% year of publication.  The corresponding reference has a \bibitem
% command in the reference list below.
%
% Please see the AASTeX manual for a more complete discussion on how to make
% \cite-\bibitem work for you.   
 
\section{Introduction}
 
Among the presently known types of carbon stars, only the CH stars have abundances and 
kinematics which are indicative of membership in the Galactic halo (McClure 1985).
Since stars of mass M $\simeq$ 0.8M$_{\odot}$ are not thought to experience the third
dredge-up mechanism during their ascent of the asymptotic giant branch (e.g., Iben 1975),
the origin of the enhanced abundances of carbon and {\sl s}-process elements in these giants
remained a puzzle until the discovery that {\sl all} field CH stars are binaries composed of 
a red-giant primary and a degenerate (i.e., white dwarf) secondary (McClure 1984; 
McClure \& Woodsworth 1990).  The peculiar abundances of these objects are therefore thought 
to be the result of mass exchange via stellar winds or Roche-Lobe overflow during the ascent 
of the white-dwarf progenitor up the asymptotic giant branch (Han et al. 1995).
Clearly, the discovery of {\sl globular cluster} CH stars would have important
implications not only for nucleogenesis and globular cluster abundance anomalies (Pilachowski et al. 1996),
but also for the formation, evolution and destruction of binaries in dense environments (Hut et al. 1992).
 
However, searches for CH stars in several globular clusters based on
spectroscopic surveys of brightest cluster members (Harding 1962), direct
imaging through intermediate-band filters (Palmer 1980; Palmer \& Wing 1982), and transmission-grating slitless
spectroscopy (Bond 1975), have proved, by and large, unsuccessful.
At present, a handful of stars having enhanced carbon and {\sl s}-process 
elements have been reported in each of $\omega$~Cen (e.g., Harding 1962; Dickens 1972; Bond 1975), 
M22 (Hesser, Hartwick \& McClure 1977; McClure \& Norris 1977; Hesser \&
Harris 1979), M55 (Smith \& Norris 1982) and M2 (Zinn 1981). 
However, while the spectra of these stars {\sl are} characterized by abnormally high CH absorption
compared to other cluster giants, they usually do {\sl not} show strong Swan bands of C$_2$,
suggesting that their anomalous carbon abundances probably arise through a different mechanism,
such as incomplete CN processing (Vanture \& Wallerstein 1992),
than that operating in ``classical" CH stars.
Indeed, among this sample of CH-enhanced stars in globular clusters, only two are likely to be 
genuine CH stars.  Both of these stars, RGO 55 (Harding 1962) and RGO 70 (Dickens 1972), are 
found in $\omega$~Cen.
 
In this {\sl Letter}, we report the serendipitous discovery of a carbon
star in the core of the poorly-studied globular cluster M14 and
argue that it is likely to be a ``classical" CH star: 
a post mass-transfer binary consisting of a red giant primary and a white dwarf 
secondary (McClure \& Woodsworth 1990). 
 
\section{Observations}
 
\subsection{MOS Spectroscopy}
 
During an observing run in May 1996 intended to measure radial velocities and Mg {\sl b} line strengths
for globular clusters surrounding the supergiant elliptical galaxy M87,
we used the MOS imaging spectrograph on the Canada-France-Hawaii telescope to obtain long-slit
comparison spectra of the integrated light of several Galactic globular clusters.
During the integrations, the telescope was allowed to drift, typically through 
a spatial extent of about one arcminute, in a direction (S $\rightarrow$ N) perpendicular to the slit in order 
to generate a spectrum representative of the integrated light of the cluster.
All spectra were accumulated using the B400 grism and STIS2 CCD 
which combine to give a dispersion of 3.6 \AA\ pixel$^{-1}$ and 
a resolution of $\simeq$ 8~\AA .  A blocking filter was used to  restrict the wavelength 
region to 4500 -- 5700 \AA .
 
Unexpectedly, the long-slit spectrum of the integrated light of the Galactic globular cluster M14,
obtained on 19 May 1996, showed strong Swan bands of C$_2$ indicating the presence of a
carbon star in the cluster core. Because of the driftscaning technique employed, the immediate
identification of the carbon star was in some doubt.  
By good fortune, however, the carbon star has few bright neighbors to
the north or south, allowing a probable identification of the candidate
based on its E-W position (see Figure 1).
On 20 May 1996, we obtained a
600s spectroscopic exposure of the suspected carbon star, this time using a 1$''$$\times$10$''$ slit. 
The blocking filter was removed to extend the spectral coverage to the region 4000 -- 8000 \AA .
After standard preprocessing, the carbon star spectrum was traced, extracted 
and wavelength calibrated using IRAF.\altaffilmark{8}\altaffiltext{8}{IRAF is distributed by the National Optical
Astronomy Observatories, which are operated by the Association of Universities for
Research in Astronomy, Inc., under contract to the National Science Foundation.}
The final spectrum is shown in Figure 2. 
 
\subsection{MOS Photometry}
 
Of the three published color-magnitude diagrams (CMDs) for M14 (e.g., Smith-Kogon,
Wehlau \& Demers 1974; Shara et al. 1986; Margon et al. 1991), only the photographic 
study of Smith-Kogon, Wehlau \& Demers (1974) includes stars brighter than the cluster
horizontal branch.
Unfortunately, these authors did not publish magnitudes, colors or 
finding charts for their program stars. Therefore, to determine 
the location of the carbon star in the cluster CMD, we obtained a number of
$B$ and $V$ images of M14 using MOS on 19 and 22 May 1996 (Figure 1).
All frames were then reduced using DAOPHOT II (Stetson et al. 1990). Empirically
we found that the CMD based on the two poorest seeing frames (FWHM $\simeq$ 1\farcs1)
exhibited less scatter, reflecting the severe MOS undersampling (0\farcs44 pixels) in the other
frames.
Our instrumental magnitudes were then calibrated using eight unsaturated local photoelectric standard stars
(Smith-Kogon, Wehlau \& Demers 1974).
These stars span a magnitude range of 12.46 $<$ $V$ $<$ 14.59
and a color range of 1.17 $<$ $B-V$ $<$ 2.05. The internal random errors in the calibration are 
$\sim$ 0.02 in $V$ and $\sim$ 0.04 in $(B-V)$.
%The location of the carbon star in the CMD is indicated by the open square in Figure 3.
 
\subsection{Hydra Spectroscopy}
 
Although the carbon star's small offset of 13\farcs8 $\simeq$ 0.3$r_c$ from the cluster 
center (Shawl \& White 1986; Trager, King \& Djorgovski 1995)
%$\simeq$ 0.3$r_c$ from the cluster center; Shawl \& White 1986; Trager, King \& Djorgovski 1995)
and its position near the tip of the cluster red giant branch (see Figure 3) suggest that it is
physically associated with the cluster, radial velocities for both the carbon star
and M14 are needed to provide unambiguous evidence of membership.\altaffilmark{9}\altaffiltext{9}{For
instance, RGO 153, a carbon star which lies only 18$'$, 
or 7$r_c$ (Trager, King \& Djorgovski 1995),
from the center of $\omega$~Cen, was shown by Smith \& Wing (1973) to have a radial velocity which 
differs from that of the cluster by more than 250 km s$^{-1}$.
For comparison, the confirmed CH stars, RGO 55 and RGO 70, are located 9$'$ and 19$'$, respectively, from the
cluster center.}
In particular, radial velocities are needed to eliminate the possibility that the object is a
carbon-rich dwarf residing in the solar neighborhood (e.g., Green et al. 1994), although the
{\sl a~priori} probability of this is low:  to a limiting magnitude of $V$ = 18, the surface density of
faint, high-latitude carbon stars (of which dwarf carbon stars make up $\sim$ 13\%) is
$\sim$ 0.02 deg$^{-2}$ (Green et al. 1994). Moreover, the MOS spectrum (Figure 2) shows
only a weak C$_2$ bandhead at 6191 \AA , a feature which Green et al. (1992)
suggest to be unusually strong in the spectra of dwarf carbon stars.
Finally, if the star in question {\sl were} a nearby dwarf, it might be expected to show a
measureable proper motion. Comparison of our CFHT images with a photographic plate
taken in 1952 with the Mt. Wilson telescope (see Figure 1 of Wehlau \& Froelich 1994) 
shows no evidence for such a proper motion.
 
On 18-19 July 1996, we used the Hydra fiber positioner and bench-mounted echelle spectrograph on the WIYN telescope to 
measure 505 radial velocities (median precision $\simeq$ 1.3 km s$^{-1}$) for 493 stars in
a $\simeq$ 0.5 deg$^{2}$ field centered on M14. 
In the course of this survey (C\^ot\'e \& Welch 1997),
we also obtained an 1800s spectroscopic exposure of the M14 carbon star at a resolution $\simeq$ 0.33 \AA\ over
the spectral range 4983 -- 5252 \AA .
After the program spectra were traced, extracted and wavelength-calibrated using IRAF, 
they were cross-correlated against a template spectrum of HD223094, a K5III radial
velocity standard.
In the case of the carbon star spectrum, the adopted template was HD156074, an R star which 
is known to have a constant heliocentric radial velocity of $-12.99$ km s$^{-1}$ (McClure 1996). 
 
\section{Discussion}
 
\subsection{Membership}
 
From our Hydra spectrum of the carbon star, we measure a radial velocity of 
$v_r = -53.0\pm1.2$ km s$^{-1}$ (heliocentric Julian date = 2450282.761).
How does this compare to the systemic velocity of M14? Available radial velocity
measurements for this cluster have been reviewed by Webbink (1981), who quotes a
weighted mean velocity of ${\bar{v_r}} = -123\pm5$ km s$^{-1}$.
This value is based primarily on low-dispersion spectra of the integrated cluster light obtained 
by Mayall (1946), as well as a small number of radial velocities (spanning the range +9 to $-$153 km s$^{-1}$) 
for four Type II Cepheids in the field of M14 accumulated by Joy (1949). An image-tube 
spectrogram obtained by Hesser, Shawl \& Meyer (1986) of the integrated cluster light
yielded a radial velocity of $v_r = -25\pm14$ km s$^{-1}$. 
Given the obvious difficulties with field star contamination in this direction
($l^{\rm II} = 21.3^{\circ}$, $b^{\rm II} = 14.8^{\circ}$), as well as
the large uncertainties of the Mayall (1946) velocities (typical internal error $\simeq$ 33 km s$^{-1}$, but
potentially much larger external errors; see Hesser, Shawl \& Meyer 1986), we have 
used our new sample of Hydra velocities to determine improved estimates of the systemic
velocity and internal dispersion of M14.\altaffilmark{10}\altaffiltext{10}{The result of 
Hesser, Shawl \& Meyer (1986) is most likely due to contamination of their spectrum
by field stars, since the distribution of radial velocities in this crowded field shows a
broad peak near $\simeq$ $-20$ km s$^{-1}$ (C\^ot\'e \& Welch 1997). 
Similarly, if the observations of Joy (1949) were made while the Type II Cepheids were near maximum light, then
the mean velocity of $-126$ km s$^{-1}$ he obtained may be (partly) explained by
atmospheric pulsations, which can approach 30--40 km s$^{-1}$ in stars of this type.}
 
We define a sample of probable M14 members by restricting ourselves to those 
stars which lie within two half-light radii $r_h$ of the cluster center (i.e., 2$r_h \simeq$ 
2\farcm6; Trager, King \& Djorgovski 1995). 
This sample consists of 20 stars which have a median velocity uncertainty of 1.2 km s$^{-1}$
and span the range $-72.0 \le v_r \le -47.6$ km s$^{-1}$. $BV$ photometry 
for these stars (see Figure 3) demonstrates that they are all likely to be cluster members since
they define a smooth red giant branch extending from $V$ = 14.38 to 16.23.
From this sample, we find a mean velocity of ${\bar{{v_r}}} = -59.5\pm1.9$ km s$^{-1}$ 
and a one dimensional cluster velocity dispersion of $\sigma_c = 8.2\pm1.4$ km s$^{-1}$ using the
maximum-likelihood estimators of Pryor \& Meylan (1993).
Based on the close agreement between the radial velocity of the carbon star and that of M14,
we conclude that the star is indeed a cluster member. We also note that, if the star
in question is similar to the field CH stars studied by McClure \& Woodsworth (1990), an additional
velocity component due to the orbital motion of the red giant primary around the center of mass 
of the system is expected; among the eight CH stars monitored by McClure \& Woodsworth (1990),
the velocity semi-amplitudes ranged from 4.3 to 12.1 km s$^{-1}$, with a mean of 8.1 km s$^{-1}$.
 
\subsection{Implications}
 
Radial velocity surveys of field CH stars have provided compelling evidence that {\sl all} of 
these systems are post mass-transfer binaries. We therefore conclude that the carbon star in M14 is very likely to be 
a spectroscopic binary having a white dwarf secondary, although {\sl confirmation} 
of duplicity must await additional radial velocity measurements.\altaffilmark{11}\altaffiltext{11}{This 
connection between the field and globular cluster CH stars has been further strengthened by the 
recent demonstration that both RGO 55 and RGO 70 in $\omega$~Cen are spectroscopic binaries 
(Mayor et al. 1996).} It is perhaps notable that, as the only two clusters known to contain 
candidate or confirmed CH stars, M14 and $\omega$~Cen are both massive, low-concentration systems.
A possible connection between cluster concentration and CH enhancement was
noted previously by McClure \& Norris (1977) who, in a prescient remark, suggested that
``searches for CH stars in the low-concentration clusters M14 and NGC 2419 might be profitable".
 
Does environment play a role in the evolution of globular cluster CH stars?
The dominant dynamical processes affecting binaries in globular clusters are the {\sl disruption} 
of ``soft" binaries through stellar encounters and the {\sl shrinking} of the orbits of ``hard" 
binaries via energy exchanges with intruder stars (e.g., Hut et al. 1992).
Equation (1) of Hills (1984) can be used to estimate the separation, $a_s$, of the widest 
cluster binaries which are expected to have escaped disruption over a Hubble time. 
Since most Galactic globular clusters have three-dimensional velocity dispersions 
$\sigma_v \lae 15$km s$^{-1}$ (Pryor \& Meylan 1993), we find $a_s$ $\gae$ 6 AU.
This is comparable to the separation of the {\sl widest} 
binaries in the McClure \& Woodsworth (1990) sample of field CH stars (i.e., 1 $\lae$ $a$ $\lae$ 4.5 AU).
Therefore, we conclude that the process of CH-star disruption is unlikely to be important for most clusters.
 
On the other hand, the shrinking of hard binaries may play a more significant role. At the cluster center,
intruder stars will 
shrink the orbits of hard binaries at the rate $${{d\ {\hbox{ln}}a}\over{dt}}\simeq{{-2\pi Ga{\rho_c}}\over{\sigma_c}}, \eqno{(1)}$$
where $a$ is the semi-major axis of the binary, ${\sigma_c}$ is the central one-dimensional velocity dispersion
and ${\rho_c}$ is the central mass density (Hills 1984; Phinney 1996).
Integration of this equation yields the initial semimajor axis $a_h$ of
the binary whose size is halved over the lifetime, $t_0$, of a cluster:
$${a_h} = {{\sigma_c} \over 2{\pi}G{\rho_c}t_0}. \eqno{(2)}$$
The orbits of hard binaries that are initially wider than this shrink
rapidly, and have $a \sim a_h$ after a Hubble time. Conversely, tighter
binaries evolve on rather longer timescales, and are largely unaffected.
Thus, $a_h$ is the characteristic final separation of surviving cluster binaries (Phinney 1996).
It is worth bearing in mind, however, that equation (2) refers to the cluster center;
the situation for the cluster as a whole is undoubtedly more complicated.
 
In Figure 4 we plot the distribution of $a_h$ against absolute magnitude, $M_V$, for all Galactic globular 
clusters (filled circles) having measured velocity dispersions and central mass densities
(e.g., see Pryor \& Meylan 1993). Absolute magnitudes are taken from Djorgovski (1993).
M14 and $\omega$~Cen are shown as open stars.
The open squares show the location of the Galactic dSph galaxies (Irwin \& Hatzidimitriou 1995). 
At least {\sl some} of the carbon stars seen in several of these galaxies (i.e., Ursa Minor,
Draco, Sculptor and Carina) are likely to be CH stars (McClure 1984; Aaronson \& Olszewski 1987; 
Armandroff, Olszewski \& Pryor 1995).
The dashed line shows the semi-major axis of the shortest-period binary
in the McClure \& Woodsworth (1990) sample of field CH stars (i.e., $a \simeq$ 1 AU). This
size is similar to the mean radii of low-mass, thermally-pulsating AGB stars 
(e.g., see Figure 2 of Boothroyd \& Sackman 1988). 
Closer initial binary orbits would disrupt the normal evolution of the 
carbon donor before it reached the thermally-pulsating AGB phase, preventing the 
dredge-up of carbon. Therefore, Figure 4 suggests that the low 
concentrations of M14 and $\omega$~Cen correspond to small values of $\rho_c/\sigma_c$, 
and hence to $a_h$ large enough to accomodate CH systems with separations typical of field 
CH stars.  
 
The discovery of a CH star in M14 therefore provides additional support for the 
notion (McClure 1984) that the process of CH star formation depends on environment, in 
particular, on the rate of hard binary shrinking through stellar encounters.
Given the incomplete and inhomogenous nature of existing surveys for CH stars
in globular clusters, a renewed effort to discover such objects in an expanded
sample of clusters (particularly those objects which lie {\sl above} the dashed line in 
Figure 4) may provide new insights into the influence of environment on binary evolution.
 
\acknowledgments
 
The authors thank the staffs of the CFH and WIYN telescopes for their outstanding support. 
Thanks also to Robert McClure and Amelia Wehlau for helpful discussions.
The research of DAH and GLHH is supported through operating grants from NSERC.

\clearpage

\figcaption[]{$B$ image of M14 taken with MOS on the CFHT. 
The carbon star is identified by the small circle, while the larger circle denotes the 
cluster core radius ($r_c$ = 50$''$ according to Trager, King \& Djorgovski 1995). 
The cross shows the location of the cluster center (Shawl \& White 1986).
North is up and east is to the 
left on this image, which measures 3$\farcm$75$\times$3$\farcm$5. \label{fig1}}
 
\figcaption[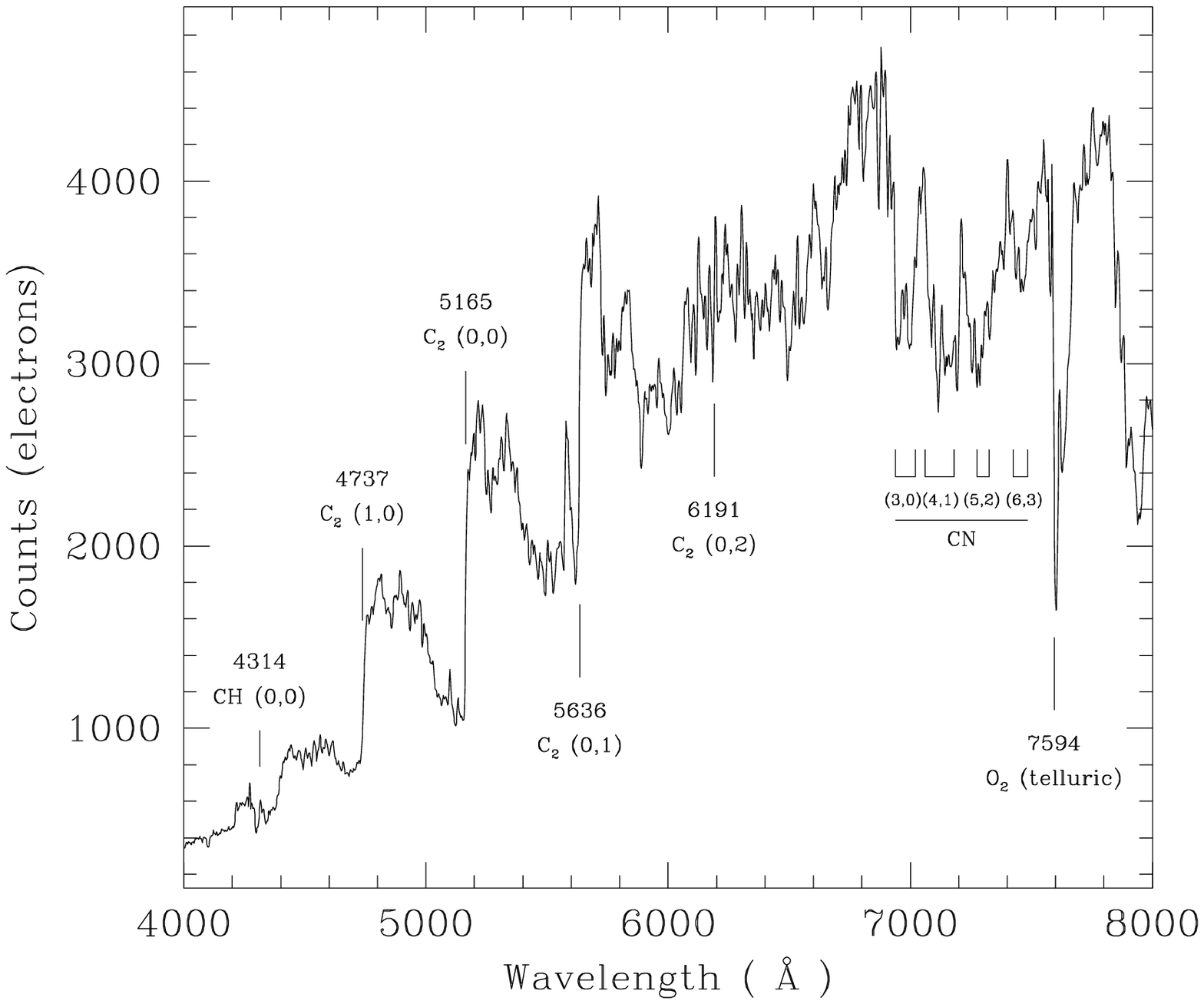]{MOS spectrum of the newly discovered M14 carbon star.
The spectrum has a dispersion of 3.6 \AA\ pixel$^{-1}$ and a resolution of $\simeq$ 8 \AA .
A number of the most prominent spectral features are indicated. \label{fig2}}
 
\figcaption[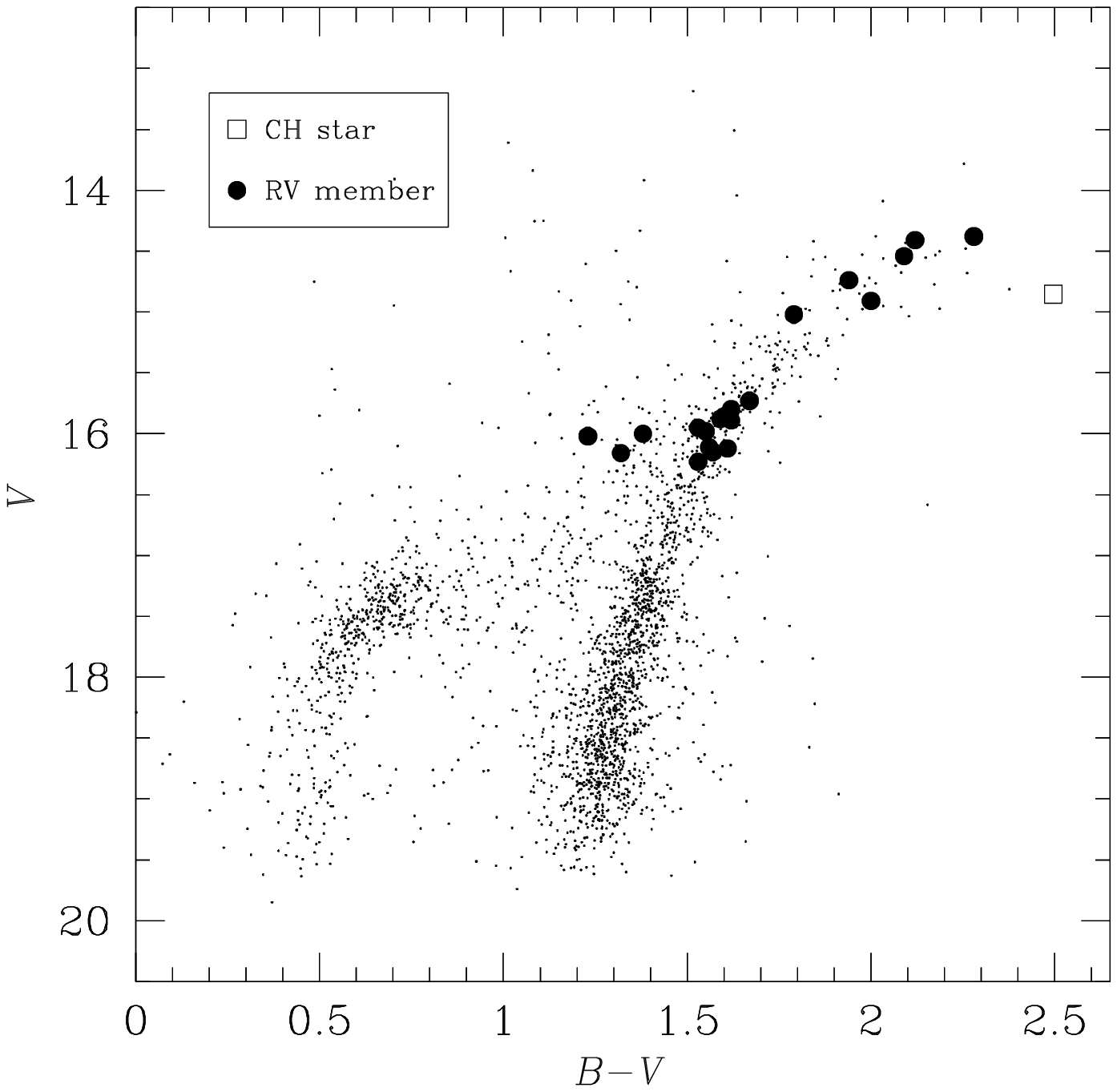]{Color-magnitude diagram (CMD) for M14 based on a pair of $BV$ images taken
with MOS. Both images have FWHM $\simeq$ 1$\farcs$1, corresponding to 2.5 pixels. 
The location of the carbon star in the CMD (i.e., $V = 14.85,~B-V = 2.50$) is indicated by the open square. 
Those stars having radial velocities measured with Hydra and lying within 2\farcm6 of the 
cluster center are indicated by the filled circles. 
\label{fig3}}
 
\figcaption[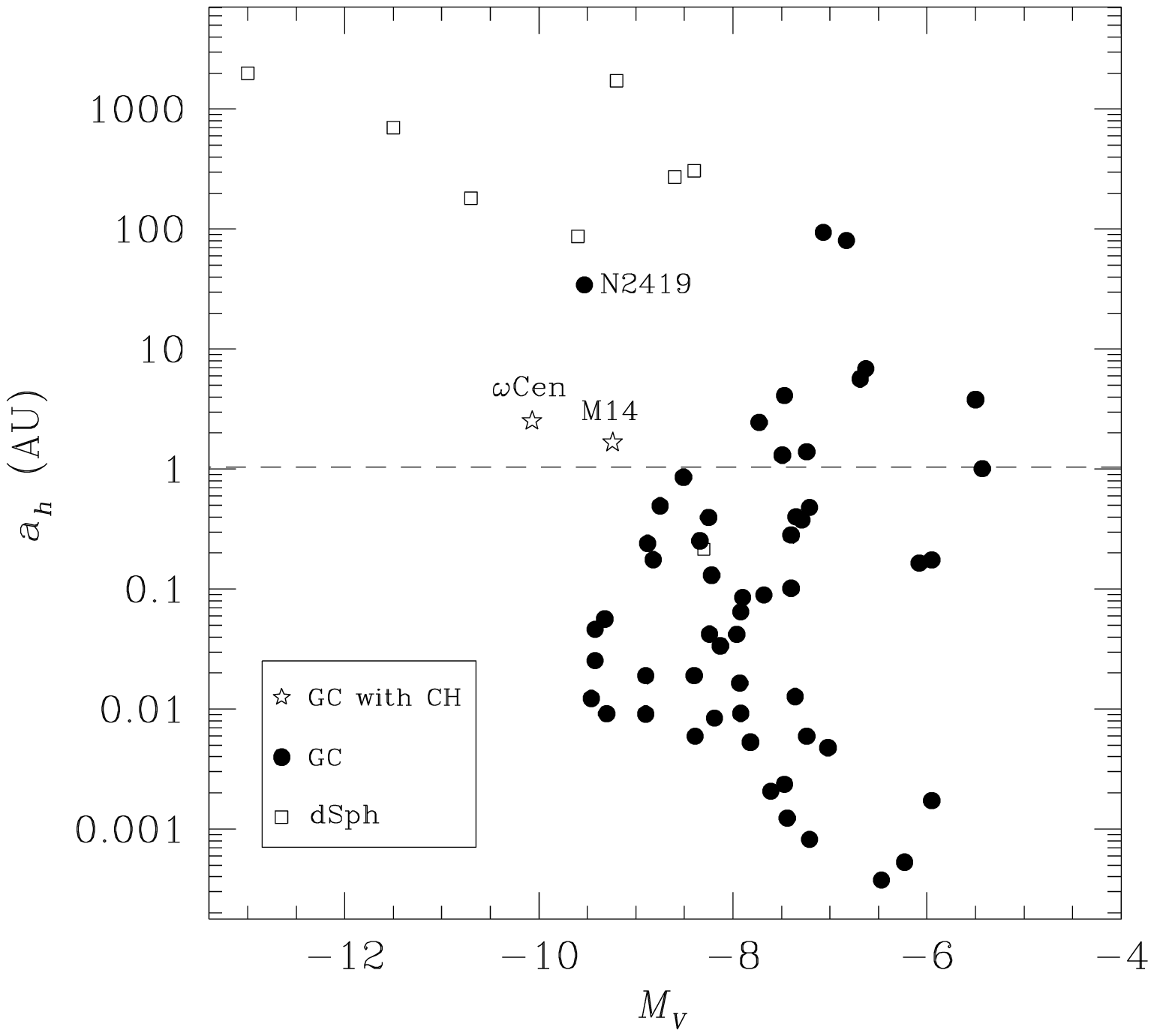]{Maximum separation of hard binaries, $a_h$, due to
shrinking via energy exchanges with intruders plotted against
absolute magnitude, $M_V$, for Galactic globular clusters (filled circles) and dSph galaxies (open squares).
The dashed line shows the semi-major axis of the shortest-period binary
in the McClure \& Woodsworth (1990) sample of field CH stars (i.e., $a$ $\simeq$ 1 AU).
M14 and $\omega$~Cen are shown as open stars. Based on its
location in this diagram, NGC 2419 is a prime candidate for a search for CH stars.\label{fig4}}
 
% Option 2.  The figure captions are printed on a caption page(s) as in 
% option 1.  The figures available as EPS files are then printed at the
% end of the document, one figure per page, using the \plotone command.
% If you wish to process this option then simply comment out the \end{document}
% just above these five lines. 
 
%\clearpage
% 
%\plotone{blank.eps}
 
\clearpage
 
\plotone{m14fig2.eps}
 
\clearpage
 
\plotone{m14fig3.eps}
 
\clearpage
 
\plotone{m14fig4.eps}
 
\end{document}